\documentclass[a4paper]{article}
\usepackage[german, english]{babel}
\usepackage{a4wide}
\usepackage{amsmath}
\usepackage{amssymb}
\usepackage{ifthen}
\usepackage{epsfig}
\newcounter{fig}

\begin{document}

\title{\bf Platonic Sphalerons}
\vspace{1.5truecm}
\author{
{\bf Burkhard Kleihaus, Jutta Kunz and Kari Myklevoll}\\
Institut f\"ur  Physik, Universit\"at Oldenburg, Postfach 2503\\
D-26111 Oldenburg, Germany}

\vspace{1.5truecm}

\date{\today}

\maketitle
\vspace{1.0truecm}

\begin{abstract}
We construct sphaleron solutions in Weinberg-Salam theory,
which possess only discrete symmetries.
Related to rational maps of degree $N$, these sphalerons
carry baryon number $Q_{\rm B}=N/2$.
The energy density of these sphalerons reflects their discrete
symmetries. We present an $N=3$ sphaleron with tetrahedral
energy density, an $N=4$ sphaleron with cubic energy density,
and an $N=5$ sphaleron with octahedral energy density.
\end{abstract}
%\vfill\eject

\section{Introduction}

As observed by 't Hooft \cite{thooft},
the standard model does not absolutely conserve
baryon and lepton number           
due to the Adler-Bell-Jackiw anomaly.
In particular 't Hooft considered spontaneous
fermion number violation due to instanton transitions
between topologically inequivalent vacua.
Manton \cite{manton} considered
the possibility of fermion number
violation in the weak interactions
from another point of view.
Showing the existence of non-contractible           
loops in configuration space, he
predicted the existence of a static, unstable solution
of the field equations, 
a sphaleron \cite{km}, representing
the top of the energy barrier between 
topologically distinct vacua.
At finite temperature the energy barrier between
distinct vacua can be overcome
due to thermal fluctuations of the fields,
and vacuum to vacuum transitions can occur,
accompanied by a change of baryon and lepton number.
The rate for baryon number violating processes
is largely determined by a Boltzmann factor, 
containing the height of the barrier at a given
temperature, and thus by the energy of the sphaleron \cite{review}.

The non-trivial topology of configuration space of Weinberg-Salam theory
gives rise to further unstable classical solutions.
A superposition of sphalerons, for instance,
leads to static axially symmetric solutions,
multisphalerons, whose energy density is torus-like \cite{kk}.
Klinkhamer, on the other hand, has constructed a static axially
symmetric solution, which may be thought of as a bound
sphaleron-antisphaleron system, in which sphaleron and
antisphaleron are located at an equilibrium distance
on the symmetry axis \cite{kl}. 
A conjectured generalization of these solutions \cite{bk}
are static axially symmetric sphaleron-antisphaleron chains \cite{kks}.

In this letter we show, that Weinberg-Salam theory possesses 
a new type of unstable classical solutions: sphalerons, which
have no rotational symmetry at all.
The symmetries of these sphalerons are only discrete,
and can be identified with the symmetries of platonic solids or crystals.
We therefore refer to them as platonic sphalerons.

Classical solutions with platonic symmetries were first observed
in the Skyrme-model of baryons and nuclei,
where these stable soliton solutions with higher baryon number
are interpreted in terms of small nuclei \cite{sk}.
Solitons with platonic symmetries are also known in the Georgi-Glashow model,
where they represent monopoles with higher magnetic charge \cite{mono},
and they arise as skyrmed monopoles in a modified Georgi-Glashow model with
higher derivative terms \cite{tigran}.

Monopoles with magnetic charge $N$ and Skyrmions with baryon number $N$
are related to rational maps of degree $N$ \cite{ratmap}.
In particular certain rational maps of degree $N$ give rise
to solitons with platonic symmetries \cite{sk,mono,tigran}.
Interestingly, the energy densities of the known classical solutions
based on the same rational map but obtained in different physical models
are qualitatively very similar.

We here base our construction of sphaleron solutions of
Weinberg-Salam theory on rational maps of degree $N$ as well.
The rational maps then determine the behaviour of the 
Higgs and gauge fields at infinity.
We solve the general set of static equations of motion numerically, 
subject to the boundary conditions specified by the rational maps.
We show, that the degree $N$ of the maps is related to the
baryon number $Q_{\rm B}$ of the sphalerons: $Q_{\rm B}=N/2$.

In particular, we here consider sphalerons based on maps 
with degree $N=1-5$. 
Besides reproducing the axially symmetric sphalerons \cite{kk},
we construct platonic sphalerons for $N=3-5$, whose energy
density has tetrahedral, cubic and octahedral symmetry, respectively.
We compare the masses of the platonic sphalerons 
with those of the axially symmetric sphalerons,
discuss the node structure of the modulus of the Higgs fields
for the platonic sphalerons, and compare with the node structure
of the corresponding platonic monopoles.
We obtain their magnetic moments perturbatively \cite{km}, 
because we construct the platonic sphalerons 
in the limit of vanishing weak mixing angle \cite{kkb,kk}.

\section{\bf Weinberg-Salam Lagrangian}

We consider the bosonic sector of Weinberg-Salam theory
\begin{equation}
{\cal L} = -\frac{1}{2} {\rm Tr} (F_{\mu\nu} F^{\mu\nu})
-  \frac{1}{4}f_{\mu \nu} f^{\mu \nu}                                           
- (D_\mu \Phi)^{\dagger} (D^\mu \Phi) 
- \lambda (\Phi^{\dagger} \Phi - \frac{v^2}{2} )^2 
\  
\label{lag1}
\end{equation}
with SU(2) field strength tensor
\begin{equation}
F_{\mu\nu}=\partial_\mu V_\nu-\partial_\nu V_\mu
            + i g [V_\mu , V_\nu ]
\ , \end{equation}
SU(2) gauge potential $V_\mu = V_\mu^a \tau_a/2$,
U(1) field strength tensor
\begin{equation}
f_{\mu\nu}=\partial_\mu A_\nu-\partial_\nu A_\mu 
\ , \end{equation}
and covariant derivative of the Higgs field
\begin{equation}
D_{\mu} \Phi = \Bigl(\partial_{\mu}
             +i g  V_{\mu} 
             +i \frac{g'}{2} A_{\mu} \Bigr)\Phi
\ , \end{equation}
where $g$ and $g'$ denote the SU(2) and U(1) gauge coupling constants,
respectively,
$\lambda$ the strength of the Higgs self-interaction and
$v$ the vacuum expectation value of the Higgs field.

The Lagrangian (\ref{lag1}) is invariant under local $SU(2)$
gauge transformations $U$,
\begin{eqnarray}
V_\mu &\longrightarrow & U V_\mu U^\dagger
+ \frac{i}{g} \partial_\mu U  U^\dagger \ ,
\nonumber\\
\Phi  &\longrightarrow & U \Phi\ .
\nonumber
\end{eqnarray}
The gauge symmetry is spontaneously broken 
due to the non-vanishing vacuum expectation
value of the Higgs field
\begin{equation}
    \langle \Phi \rangle = \frac{v}{\sqrt2}
    \left( \begin{array}{c} 0\\1  \end{array} \right)   
\ , \end{equation}
leading to the boson masses
\begin{equation}
    M_W = \frac{1}{2} g v \ , \ \ \ \ 
    M_Z = \frac{1}{2} \sqrt{(g^2+g'^2)} v \ , \ \ \ \ 
    M_H = v \sqrt{2 \lambda} \ . 
\end{equation}
$ \tan \theta_w = g'/g $ determines
the weak mixing angle $\theta_w$,
defining the electric charge $e = g \sin \theta_w$.  

In Weinberg-Salam theory, baryon number is not conserved
\begin{equation}
 \frac{d Q_{\rm B}}{dt} = \int d^3 r \partial_t j^0_{\rm B}
= \int d^3 r \left[ \vec \nabla \cdot \vec j_{\rm B}
 + \frac{g^2}{32 \pi^2} \epsilon^{\mu\nu\rho\sigma} \,
{\rm Tr} \left(F_{\mu\nu} F_{\rho\sigma} \right) \right] \ . 
\end{equation}
Starting at time $t=-\infty$ at the vacuum with $Q_{\rm B}=0$,
one obtains the baryon number of a sphaleron solution at
time $t=t_0$ \cite{km},
\begin{equation}
 Q_{\rm B} = 
\int_{-\infty}^{t_0} dt \int_S \vec K \cdot d \vec S
+  \int_{t=t_0} d^3r K^0 \ , 
\end{equation}
where the $\vec \nabla \cdot \vec j_{\rm B}$ term is neglected,
and the anomaly term is reexpressed in terms of the
Chern-Simons current
\begin{equation}
 K^\mu=\frac{g^2}{16\pi^2}\varepsilon^{\mu\nu\rho\sigma} {\rm Tr}(
 F_{\nu\rho}V_\sigma
 + \frac{2}{3} i g V_\nu V_\rho V_\sigma )
\ . \end{equation}
In a gauge, where
\begin{equation}
V_\mu \to \frac{i}{g} \partial_\mu \hat{U} \hat{U}^\dagger \ , \ \ \ 
\hat{U}(\infty) = 1 \ , 
\end{equation}
$\vec K$ vanishes at infinity, yielding for the baryon charge
of a sphaleron solution
\begin{equation}
 Q_{\rm B} = \int_{t=t_0} d^3r K^0 \ .
\label{Q}
\end{equation}

Here we are interested in static classical solutions of the general
field equations with vanishing time components of
the gauge fields, $V_0=0$ and $A_0=0$.
For non-vanishing $g'$ it is inconsistent to set the U(1) field to
zero, since the U(1) current 
\begin{equation}
j_i = -\frac{i}{2} g' (\Phi^\dagger D_i \Phi - (D_i\Phi)^\dagger \Phi) \ 
\label{u1current}
\end{equation}
acts as a source for the gauge potential $A_i$.
This current also determines the
magnetic moment $\vec{\mu}$ of a classical configuration, since
\begin{equation}
\vec{\mu} = \frac{1}{2} \int \vec{r} \times \vec{j} d^3r \ .
\label{mu}
\end{equation}
When $g'=0$, the U(1) gauge potential $A_\mu$ decouples
and may consistently be set to zero.
Since we here construct sphaleron solutions 
in the limit of vanishing Weinberg angle, 
we determine their magnetic moments only perturbatively \cite{km}.
We note, that the ratio $\vec{\mu}/e$ remains finite 
for $\theta_w \rightarrow 0$.

\section{Rational maps}

To obtain sphaleron solutions with discrete symmetry
we make use of rational maps,
i.e.~holomorphic functions from $S^2\mapsto S^2$ \cite{ratmap}.
Treating each $S^2$ as a Riemann sphere, the first having coordinate 
$\xi$,
a rational map of degree $N$ is a function $R:S^2\mapsto S^2$ where
\begin{equation}
R(\xi)=\frac{p(\xi)}{q(\xi)} 
\ , \label{rat} \end{equation}
and $p$ and $q$ are polynomials of degree at most $N$, where at least
one of $p$ and $q$ must have degree precisely $N$, and $p$ and $q$
must have no common factors \cite{ratmap}.

We recall that via stereographic projection, the complex coordinate $\xi$
on a sphere can be identified with conventional polar coordinates by
$\xi=\tan(\theta/2)e^{i\varphi}$ \cite{ratmap}.
Thus the point $\xi$ corresponds to the unit vector
\begin{equation}
\vec {n}_\xi=\frac{1}{1+\vert \xi \vert^2}
(2\Re(\xi), 2\Im(\xi),1-\vert \xi \vert^2)
\ , \label{unit1} \end{equation}
and the value of the rational map $R(\xi)$ 
is associated with the unit vector
\begin{equation}
\vec {n}_R=\frac{1}{1+\vert R \vert^2}
(2\Re(R), 2\Im(R),1-\vert R\vert^2).
\label{unit2}
\end{equation}
 
Parametrizing the Higgs field as 
\begin{equation}
\Phi = (\Phi_0 1\hspace{-0.28cm}\perp + i \Phi_a \tau_a)\frac{v}{\sqrt2}
    \left( \begin{array}{c} 0\\1  \end{array} \right) \ ,
\end{equation}
we impose at infinity the boundary conditions
\begin{equation}
\Phi_0 = 0 \ , \ \ \ 
\Phi_a  \tau_a=  (\vec {n}_R)\cdot {\vec \tau} =: \tau_R
\ . \label{bcHiggs} \end{equation}
The boundary conditions for the gauge field are obtained from 
the requirement $D_i \Phi =0$ at infinity, yielding
\begin{equation}
V_i = \frac{i}{g} (\partial_i \tau_R )\tau_R
\ , \label{bcA} \end{equation}
i.~e.~the gauge field tends to a pure gauge at infinity, 
$V_i = \frac{i}{g} (\partial_i U_\infty) U_\infty^\dagger $, 
with $U_\infty=i \tau_R$.

Subject to these boundary conditions,
and the gauge condition
\begin{equation}
\partial_i V^i =0
\ , \label{gaugecond} \end{equation}
we then solve the general set of field equations, involving
4 functions $\Phi_0(x,y,z)$, $\Phi_a(x,y,z)$ for the Higgs field 
and 9 functions $V_i^a(x,y,z)$ for the gauge field,
and $V_0^a=0$.
The solutions additionally satisfy the condition
$\partial^\mu {\rm Tr}\, ( V_\mu \Phi) =0$,
corresponding to $\Phi_0(x,y,z)=0$.

We here consider platonic sphalerons obtained from maps $R_N$,
\begin{equation}
R_3(\xi)=\frac{\sqrt{3}a\xi^2-1}{\xi(\xi^2-\sqrt{3}a)} \ , \ \ a=\pm i \ , \
\label{map1} \end{equation}
\begin{equation}
R_4(\xi)=c\frac{\xi^4+2\sqrt{3}i\xi^2+1}{\xi^4-2\sqrt{3}i\xi^2+1} \ , \ \ c=1 \ , \
\label{map2} \end{equation}
\begin{equation}
R_5(\xi)=\frac{\xi(\xi^4+b\xi^2+a)}{a\xi^4-b\xi^2+1} \ , \ \ b=0\ ,\ a=-5  \ . \
\label{map3} \end{equation}
Note, that the choice $a=0$ in (\ref{map1}),
and $a=b=0$ in (\ref{map3}) yields the axially symmetric
sphalerons of ref.~\cite{kk} for $N=3$ and $N=5$, respectively, 
in a different gauge,
while the axially symmetric sphaleron for $N=4$ is obtained from 
$R_4(\xi)= \xi^4$.

The baryon number $Q_{\rm B}$ of the sphalerons is obtained from
Eq.~(\ref{Q}), after performing a gauge transformation with
\begin{equation}
U = \exp(-i \Omega(x,y,z) \tau_R/2) \ , 
\label{gtu} \end{equation}
where $\Omega$ tends to $\pi$ at infinity and vanishes at
the origin.
We note, that the non-gauge transformed Chern-Simons density $K^0$ 
vanishes identically for the spherically and axially symmetric sphalerons, 
due to the ansatz of the gauge potential \cite{foot1}. 
In contrast,
for the platonic sphalerons the non-gauge transformed 
Chern-Simons density $K^0$ is non-trivial,
and we checked numerically that it does not contribute to the 
baryon number for the platonic sphalerons.
Thus the only contribution arises from the gauge transformation $U$,
Eq.~(\ref{gtu}).
Consequently, the platonic sphalerons have baryon number 
\begin{equation}
Q_{\rm B} = \frac{N}{2} \ .
\end{equation}

\section{\bf Platonic Sphalerons}

To construct platonic sphalerons,
we transform to dimensionless coordinates
$\tilde{x} = xvg\ , \ \tilde{y} = yvg\ , \tilde{z} = zvg$, and scale the 
gauge potential $V_\mu \to v V_\mu$.
The set of classical equations of motion is then solved numerically,
subject to the boundary conditions specified by
(\ref{bcHiggs})-(\ref{bcA}) for the rational maps 
(\ref{map1})-(\ref{map3}).
We employ the Gau\ss-Seidel algorithm on an equidistant mesh
in the coordinates $(\bar x, \bar y,\bar z)$ defined by
\begin{equation}
\tilde{x} =
R_{\rm L} \frac{\sin\bar x}{\cos^2\bar x}\frac{\cos^2 \alpha}{\sin\alpha}
 \ , \ \ \
\tilde{y} =
R_{\rm L} \frac{\sin\bar y}{\cos^2\bar y}\frac{\cos^2 \alpha}{\sin\alpha}
 \ , \ \ \
\tilde{z} =
R_{\rm L} \frac{\sin\bar z}{\cos^2\bar z}\frac{\cos^2 \alpha}{\sin\alpha}  
 \ , \ \ \
\label{map} \end{equation}
where $R_{\rm L}$ defines the extend of the integration volume 
and $\alpha < \pi/2$ defines the range $[-\alpha,\alpha]$ 
of the coordinates $\bar x$, $\bar y$, $\bar z$. 
The numerical solutions are obtained with $\alpha=1.082$ and
$R_{\rm L}=12$ for $N=3$ and $R_{\rm L}=15$ for $N=4,5$.
The mesh consists of $71$ meshpoints in each direction.

The numerical solutions satisfy 
the relation among the energy contributions
\begin{equation}
\int \frac{1}{2} {\rm Tr} (F_{\mu\nu} F^{\mu\nu}) d^3r
= \int (D_\mu \Phi)^{\dagger} (D^\mu \Phi)  d^3r
+3 \lambda \int (\Phi^{\dagger} \Phi - \frac{v^2}{2} )^2  d^3r
\ , \end{equation}
obtained from a scaling argument,
with an accuracy of $10^{-2}$.
$\Phi_0(x,y,z)=0$ within the numerical accuracy.

Turning to the numerical results,
we first address the energy density of the platonic sphalerons.
Defining the energy density $\varepsilon$ by
\begin{equation}
M= \frac{1}{4\pi} \int \varepsilon (\vec x)
                       dx dy dz
\ , \end{equation}
where $M$ is the mass in units of $4 \pi v/g$, 
we present surfaces of constant energy density $\varepsilon$ in Figs.~1
for the platonic sphalerons
based on the maps (\ref{map1}), (\ref{map2}), and (\ref{map3}), for $M_H=M_W$.
%(i.e.~$\lambda=g^2/2$).
The energy density of these sphalerons
clearly exhibits tetrahedral, cubic and octahedral symmetry, respectively.
Since the energy density of a platonic sphaleron is qualitatively 
very similar to the energy density
of a platonic monopole and a platonic Skyrmion obtained from the same map
\cite{ratmap},
this indicates, that the shape of the energy density of a classical solution
is determined primarily by the rational map, 
and rather independent of the model and the stability of the solution.

In Table~1 we present the masses of these platonic sphalerons
with $N=3-5$ in units of $4 \pi v/g$ 
for two values of the Higgs mass,
$M_H=M_W$ and $M_H=2 M_W$.
Also exhibited are the masses of the axially symmetric sphalerons \cite{kk}
with winding number $N=2-5$, $N=1$ represents the spherically
symmetric sphaleron \cite{manton}.
For the Higgs masses considered,
the mass $M(N)$ of the platonic sphalerons is slightly smaller than
the mass of the corresponding axially symmetric sphalerons.
(We note, that the mass difference is significantly larger than
the numerical error for the masses.)
Likewise, the mass of platonic Skyrmions is smaller than the mass of
the corresponding axially symmetric Skyrmions \cite{sk}.
In contrast, the mass of platonic skyrmed monopoles 
is slightly higher than the mass of the corresponding
axially symmetric skyrmed monopoles \cite{tigran}.
Comparing the mass $M(N)$ of the platonic sphalerons
with $N$ times the mass $M(1)$ of the spherically symmetric sphaleron,
we observe, that for the Higgs masses considered here, 
their ratio $M(N)/N\, M(1)$ is close to one.
(For axially symmetric sphalerons, the mass ratio $M(N)/N\, M(1)$
is smaller than one for small Higgs masses and larger than one for large
Higgs masses \cite{kk}.)

We next address the modulus of the Higgs field,
and in particular, the location of its nodes.
All sphalerons possess a node at the origin, which is the only node
for the spherically symmetric sphaleron \cite{manton}
and the axially symmetric sphalerons \cite{kk}.
For the platonic sphalerons we expect a pattern of nodes,
in accordance with the symmetries of the solutions.
For the tetrahedral sphaleron ($N=3$), for instance,
four ($N+1$) additional nodes may be located along the four spatial diagonals,
which pass through the maxima of the energy density.
Alternatively, four additional nodes could be located at the centers of
the faces of the tetrahedron.

In Figs.~2 we exhibit 
the components of the Higgs field $\Phi_a$, $a=1,2,3$,
in units of $v/\sqrt{2}$ along those spatial directions,
where in accordance with the symmetries of the platonic sphalerons
nodes of the modulus of the Higgs field may be found. 
At a node of the modulus of the Higgs field
all components must vanish.
As seen in Figs.~2a,b, the modulus of the Higgs field of
the tetrahedral sphaleron has indeed five nodes,
four located on the diagonals close to the maxima of the energy density, 
and one located at the origin.
One may thus be tempted to interpret the tetrahedral sphaleron
as a superposition of four sphalerons ($N=1$) located at the nodes
along spatial diagonals and one antisphaleron ($N=-1$) located at the origin.
The energy density at the origin is small however.

Similarly, we observe seven nodes for the modulus of the Higgs field
of the octahedral sphaleron ($N=5$), as seen in Fig.~2c. 
In accordance with the symmetries, one node is located at the origin, 
and six ($N+1$) nodes are located symmetrically on the cartesian axes. 
Again these six nodes are associated with 
the six maxima of the energy density, 
suggesting to interpret the octahedral sphaleron
as a superposition of six sphalerons ($N=1$) and one antisphaleron ($N=-1$),
although the energy density at the central node is small.
In contrast to the $N+2$ nodes of the tetrahedral and octahedral sphalerons,
we observe only a single node for the cubic sphaleron ($N=4$).
There are no nodes along the spatial diagonals 
close to the eight ($N+3$) maxima of the energy density, 
which would be required for an interpretation of the cubic sphaleron
in terms of a superposition of sphalerons ($N=1$) and antisphalerons,
and there are no nodes along the cartesian axes as well.

Comparing the node structure of the platonic sphalerons with the
node structure of the platonic monopoles,
we note, that they are completely analogous for the rational maps
considered \cite{ratmap}.
Associating for the monopoles a node of the Higgs modulus 
with the location of a magnetic charge, 
an $N=3$ tetrahedral monopole, for instance, would then be composed
of four monopoles and one antimonopole, and thus possess
the proper total magnetic charge \cite{zeros}.

In Table~1 we also exhibit the magnetic moment $\mu$
in units of $2\pi g'/(3 g^3 v)$ of the platonic 
and axially symmetric sphalerons \cite{kk}.
The magnetic moment is obtained perturbatively, since
the sphaleron solutions are constructed in the limit
of vanishing weak mixing angle.
(For the axially symmetric sphalerons
the deviation of the perturbative value of $\mu$ 
from the non-perturbative value \cite{kk}
is only 1\% for $\theta_w=0.5$ and $M_H=M_W$.)
The magnetic moment $\mu(N)$ of the axially symmetric sphalerons
increases strongly with $N$,
yielding a ratio $\mu(N)/N\mu(1)$ on the order of one.
The magnetic moment $\mu(N)$ of the corresponding platonic sphalerons
is considerably smaller, yielding a ratio
$\mu(N)/N\mu(1)$ of only about one third for $N=3$ and $N=5$,
while the magnetic moment of the $N=4$ platonic sphaleron
vanishes (within numerical accuracy: $\mu < 10^{-3}$).

\section{\bf Conclusions}

We have constructed sphalerons in Weinberg-Salam theory
which possess only discrete symmetries.
These sphalerons are based on rational maps of degree $N$
and have baryon number $Q_{\rm B}=N/2$.
The energy density of the platonic sphalerons constructed for $N=3-5$
possesses tetrahedral, cubic and octahedral symmetry.
Interestingly,
the energy densities of platonic sphalerons are qualitatively
very similar to the energy densities
of platonic monopoles and platonic Skyrmions, 
obtained from the same rational map \cite{ratmap,tigran}.
We thus conclude, that the shape of the energy density of a classical solution
is determined primarily by the rational map,
and rather independent of the model and the stability of the solution.

The mass $M(N)$ of the platonic sphalerons is lower than the mass
of the corresponding axially symmetric sphalerons,
for the Higgs masses considered,
and their mass ratio $M(N)/N\, M(1)$ is close to one.
For Skyrmions, monopoles and skyrmed monopoles
a second map with degree $N=5$ has been considered,
leading to solutions with dihedral symmetry \cite{ratmap,tigran}.
The dihedral Skyrmion and skyrmed monopole possess a
slightly smaller mass than their octahedral counterparts \cite{ratmap,tigran}.
We expect a dihedral sphaleron ($N=5$) in Weinberg-Salam theory as well,
and also platonic sphalerons based on maps of higher degree $N>5$.

The node structure of the modulus of the Higgs field
of the platonic sphalerons and of the platonic monopoles
is also completely analogous for the rational maps considered.
The tetrahedral sphaleron has five nodes,
four located on the diagonals close to the maxima of the energy density,
and one located at the origin.
Similarly, the octahedral sphaleron has seven nodes,
six located symmetrically on the cartesian axes 
close to the maxima of the energy density, and one located at the origin.
The cubic sphaleron, in contrast, has a single node located at the origin.

The perturbatively obtained
magnetic moments of the platonic sphalerons don't exhibit
the (almost) linear growth with $N$,
observed for the axially symmetric sphalerons \cite{kk}.
The magnetic moments of the tetrahedral and octahedral sphalerons
are only about one third of the magnetic moments of the corresponding
axially symmetric sphalerons, and the magnetic moment of the cubic
sphaleron vanishes.
The influence of a finite mixing angle on the magnetic moment
and on the masses is expected to be small \cite{kk},
and will be considered elsewhere.

The solutions constructed here possess an additional property,
they satisfy $\partial^\mu {\rm Tr} V_\mu \Phi =0$,
corresponding to $\Phi_0(x,y,z)=0$.
Thus only three of the four Higgs field functions are non-trivial.
Without this symmetry property more general solutions may be found.
In the case of spherical symmetry, for instance,
additional unstable solutions, bisphalerons, appear \cite{kb}.
Their generalization to axial symmetry and platonic symmetries
remains open.

We conclude from our results,
that the occurrence of localized finite energy solutions
with platonic symmetries is a more general phenomenon than
previously thought, since it appears to be present in various
non-Abelian field theories. 
In particular, since the solutions constructed here are sphalerons, 
stability is clearly not needed for such solutions to exist.
Consequently, we expect the presence of platonic sphalerons 
in further theories, such as
pure Yang-Mills theory coupled to gravity \cite{kkreg}.

Acknowledgement: 
B.K.~gratefully acknowledges support by the DFG under contract
KU612/9-1, and K.M.~by the Research Council of Norway under 
contract 153589/432.

%\vfill\eject

%\newpage

\vspace{3 cm}
{\bf Table 1} \\
\begin{center}
\vspace{3 mm}
$ M(N)\ [4\pi v/g]$   $(M(N)/N\, M(1)) $ and
$ \mu(N) [2\pi g'/(3 g^3 v)]$   $(\mu(N)/N\, \mu(1)) $ \\
\vspace{3 mm}
\begin{tabular}{||c|c|c|c|c||} \hline\hline
$N$   & $M_H =  M_W$ & $ M_H = 2 M_W$ & 
        $M_H =  M_W$ & $ M_H = 2 M_W$ \\
\hline
$ 1 $   & $ 1.82$ $(1.00)$       & $1.98   $   $(1.00)$ 
& $ 21.12$  $(1.00)$             & $ 19.18 $   $(1.00)$ \\
$ 2^* $ & $ 3.60$ $(0.99)$       & $4.03   $   $(1.02)$ 
& $ 44.3\ $ $(1.05)$             & $ 38.9\ $   $(1.01)$ \\
$ 3 $   & $ 5.33$ $(0.98)$       & $6.09   $   $(1.03)$ 
& $24.1\ $  $(0.38)$             & $ 22.0\ $   $(0.38)$ \\
$ 3^* $ & $ 5.44$ $(1.00)$       & $6.19   $   $(1.04)$ 
& $ 70.7\ $ $(1.12)$             & $ 60.9\ $   $(1.06)$ \\
$ 4 $   & $ 7.07$ $(0.97)$       & $8.19   $   $(1.04)$ 
& $ 0. \ $  $(0.)   $            & $0.\ \  $   $(0.)$ \\
$ 4^* $ & $ 7.34$ $(1.01)$       & $8.46   $   $(1.07)$ 
& $ 100.2$  $(1.19)$             & $ 84.2\ $   $(1.10)$ \\
$ 5 $   & $ 8.90$ $(0.98)$       & $10.36  $   $(1.05)$ 
& $39.0\ $  $(0.37) $            & $ 35.2\ $   $(0.37)$ \\
$ 5^* $ & $ 9.30$ $(1.02)$       & $10.83  $   $(1.10)$ 
& $132.2$  $(1.25)$              & $ 110.0 $   $(1.15)$ \\
\hline
\hline
\end{tabular}

\end{center}

%\vspace{1cm}

The masses $M(N)$ of sphaleron solutions based on maps of degree $N=1-5$
are presented in units of $4\pi v/g$ for $M_H=M_W$ and $M_H=2 M_W$,
together with the mass ratios $(M(N)/N\, M(1))$.
$N^*$ configurations represent axially symmetric sphalerons.
Also shown are the magnetic moments $\mu$ in units of $2\pi g'/(3 g^3 v)$
and the magnetic moment ratios $(\mu(N)/N\, \mu(1)) $.

\newpage

{\bf Figure 1:}

%\mbox{\epsfysize=9.0cm \epsffile{../../../kleihaus/mdis/pap/eN3_mw_0.4.ps} }

\parbox{\textwidth}{
\centerline{
\mbox{\epsfysize=6.0cm \epsffile{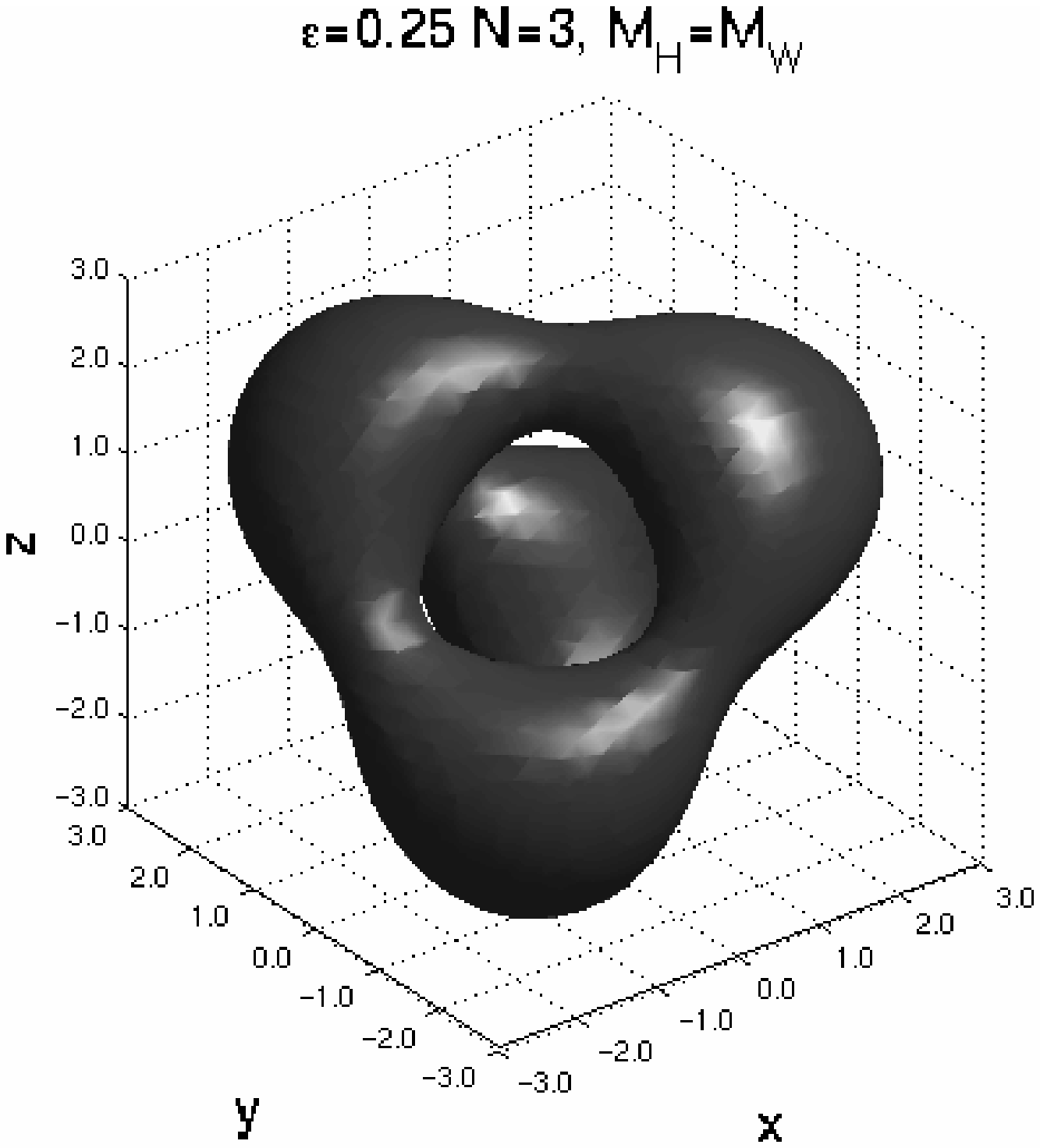} } \hspace{1cm}
\mbox{\epsfysize=6.0cm \epsffile{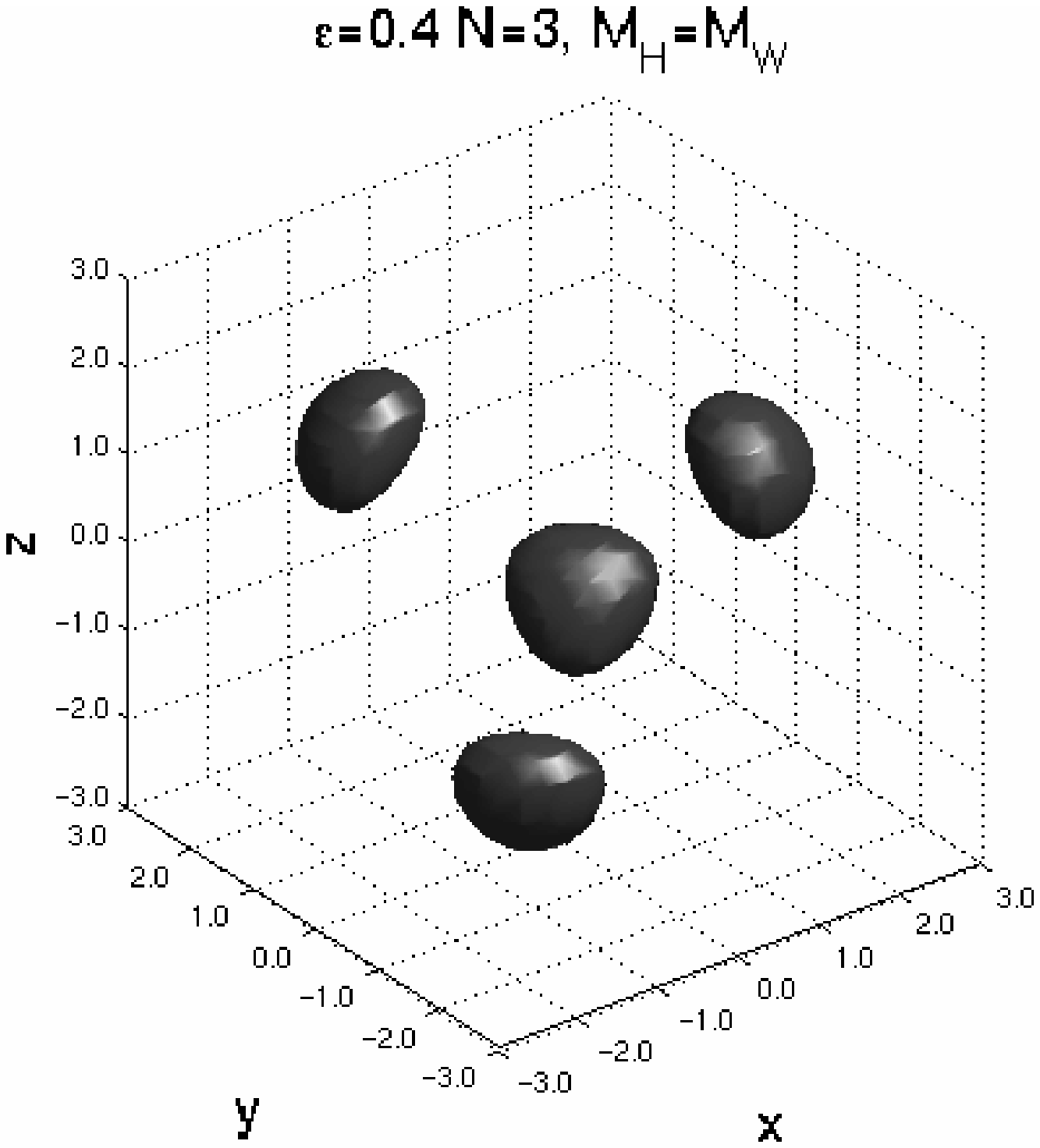} }
}\vspace{1.cm} }
\parbox{\textwidth}{
\centerline{
\mbox{\epsfysize=6.0cm \epsffile{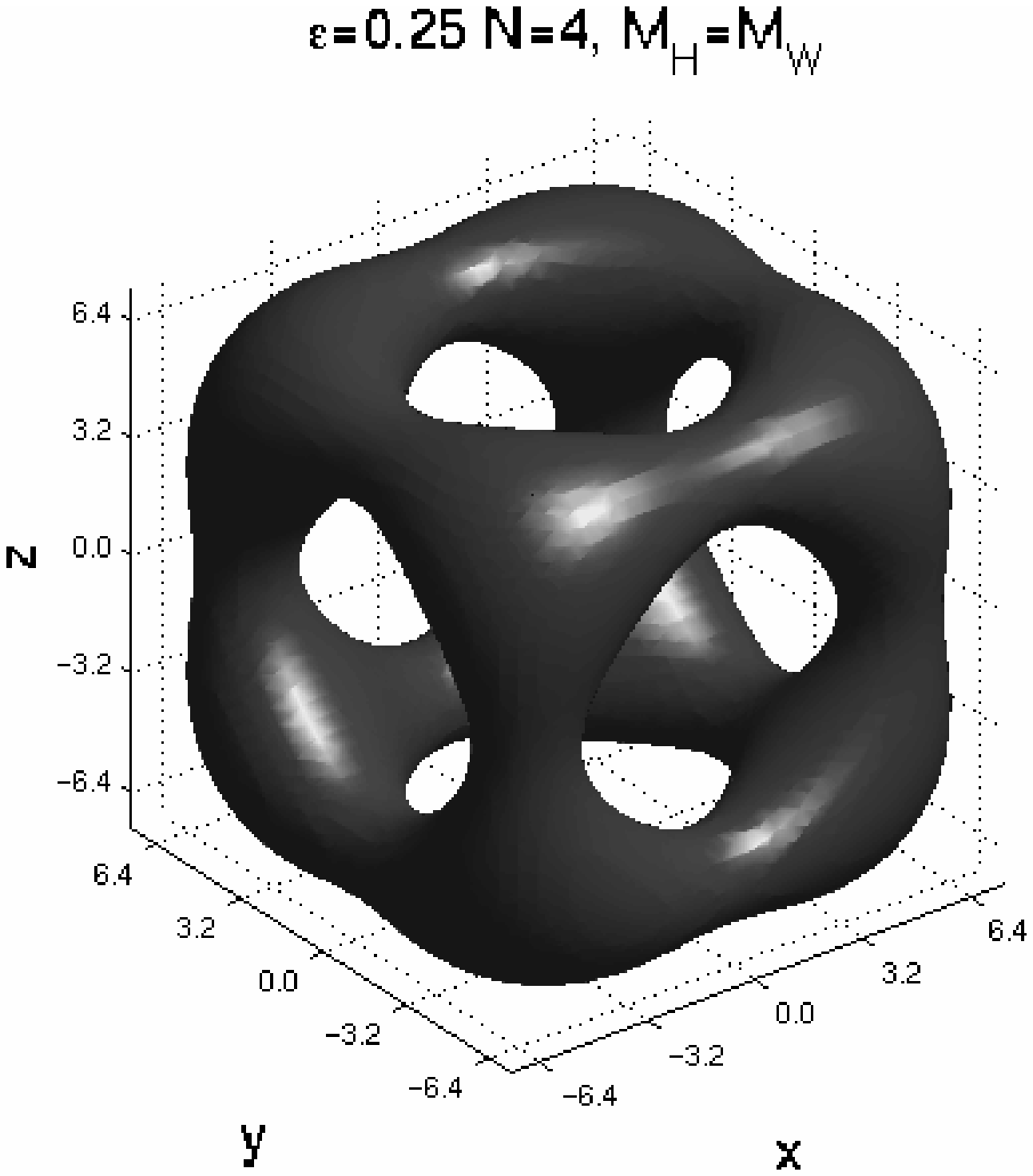} } \hspace{1cm}
\mbox{\epsfysize=6.0cm \epsffile{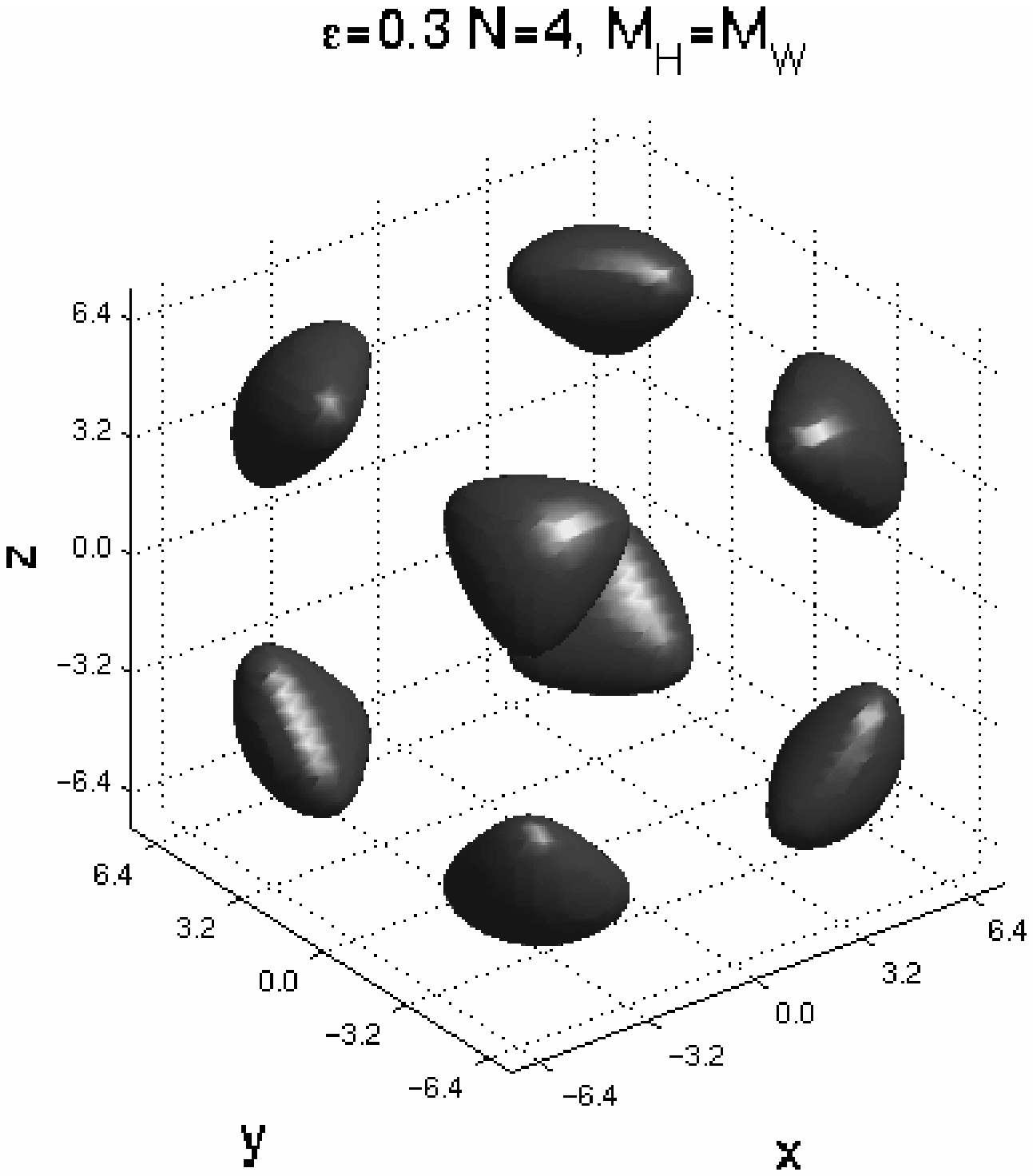} }
}\vspace{1.cm} }
\parbox{\textwidth}{
\centerline{
\mbox{\epsfysize=6.0cm \epsffile{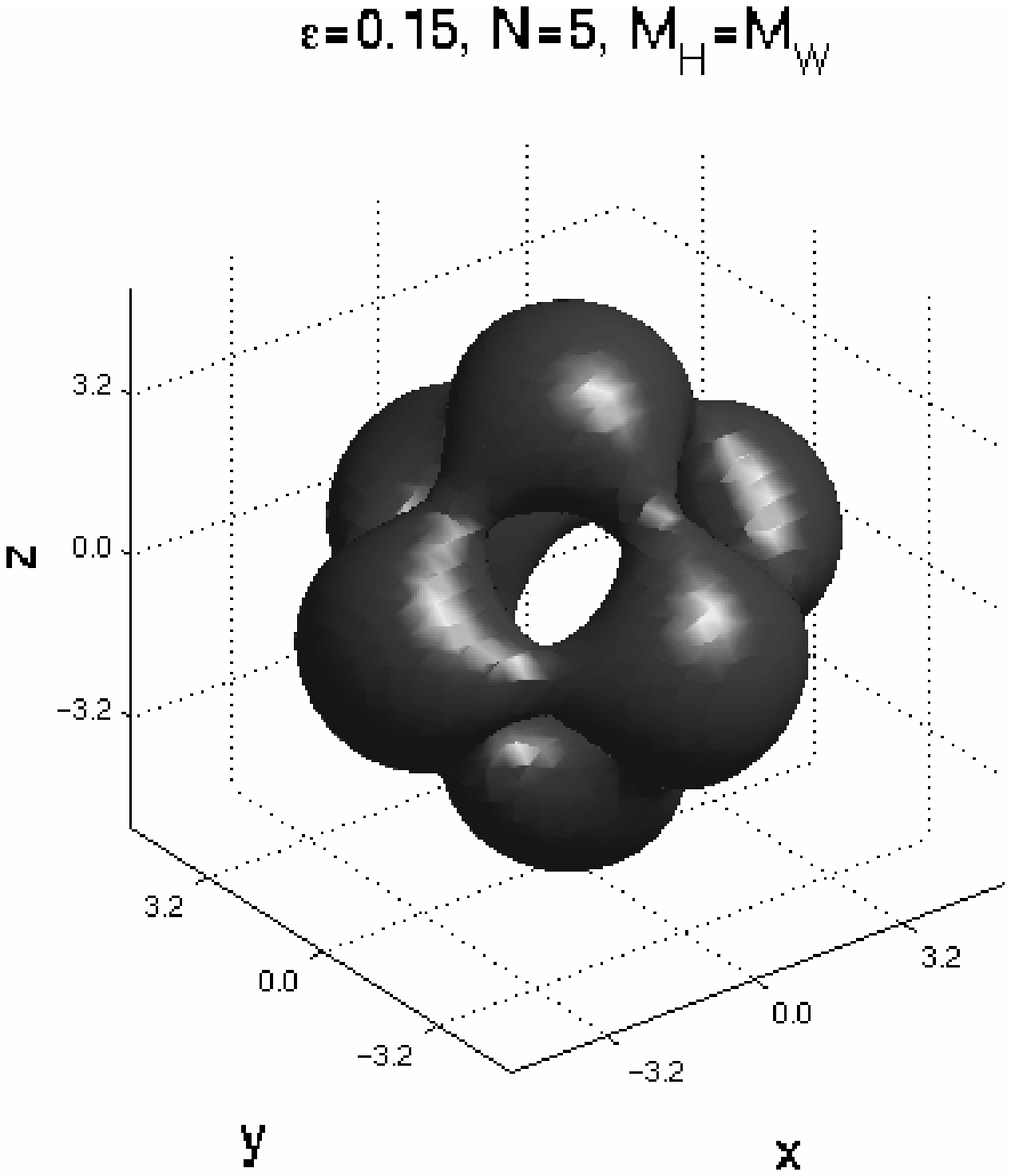} } \hspace{1cm}
\mbox{\epsfysize=6.0cm \epsffile{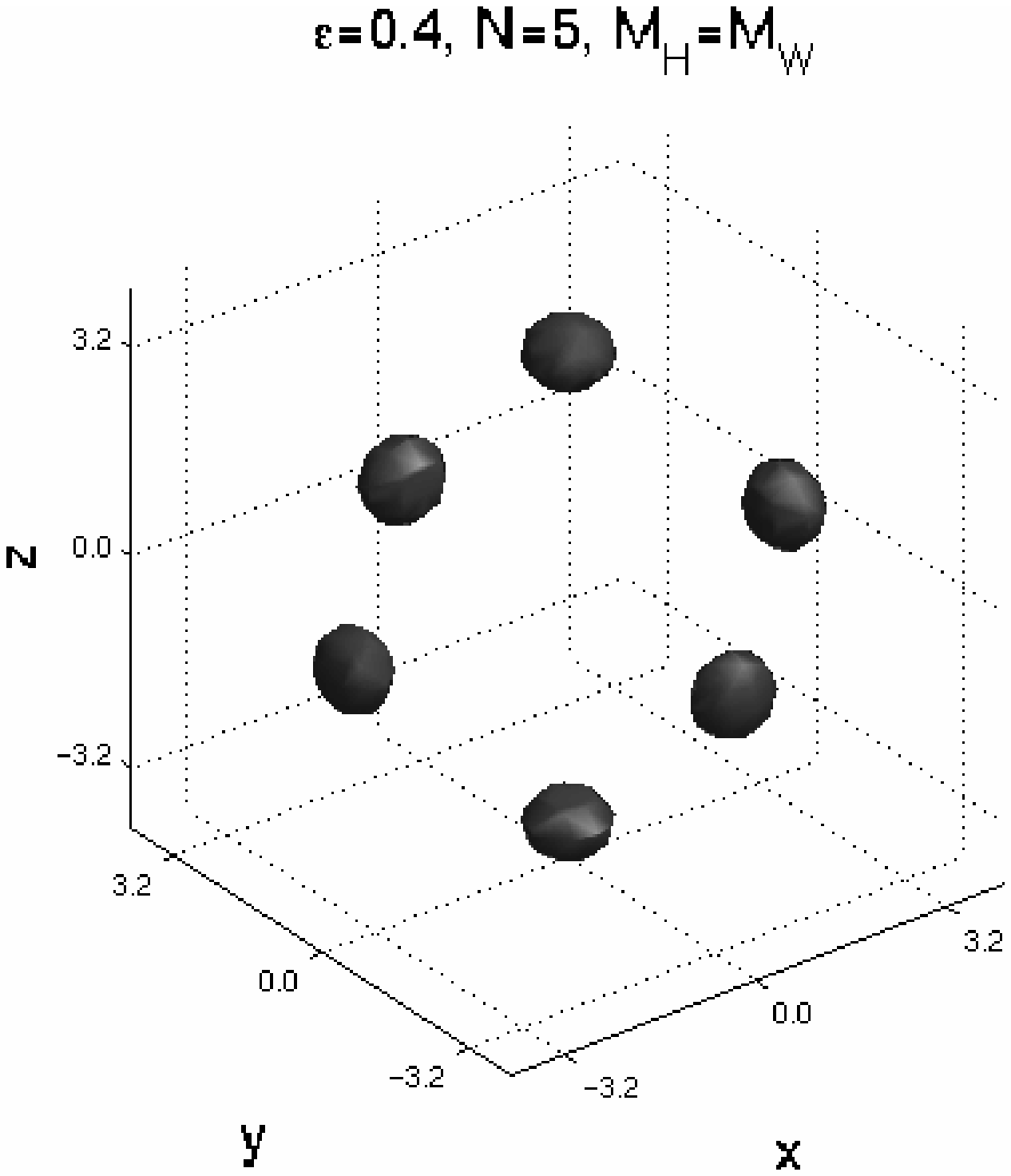} }
}\vspace{1.cm} }

Surfaces of constant energy density $\varepsilon$ are shown 
for the tetrahedral sphaleron ($N=3$), 
the cubic sphaleron ($N=4$), 
and the octahedral sphaleron ($N=5$).

\newpage

{\bf Figure 2:}

\noindent
\parbox{\textwidth}{
\centerline{
\mbox{\epsfysize=6.0cm 
\epsffile{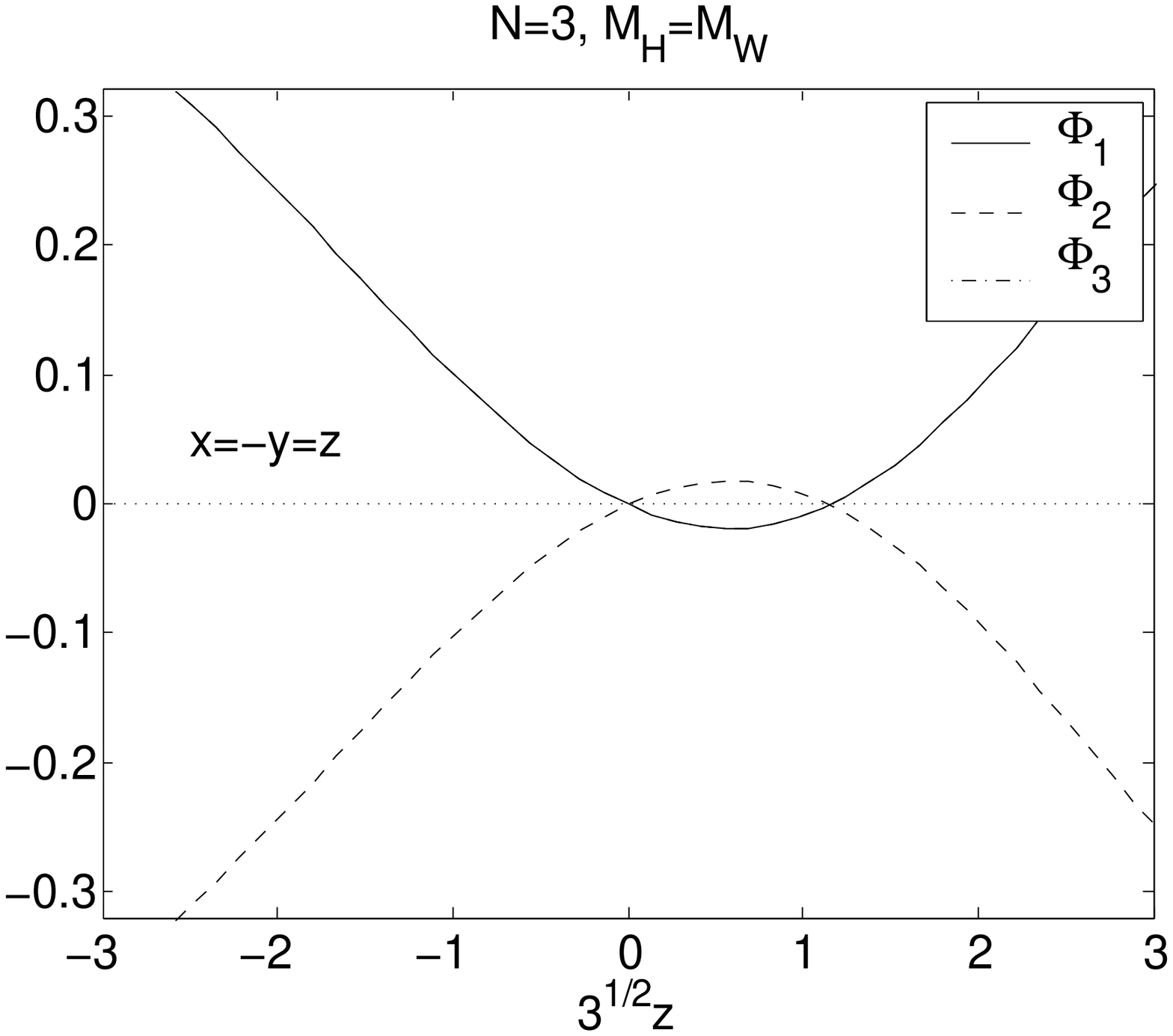} \hspace{1.cm}
\epsfysize=6.0cm
\epsffile{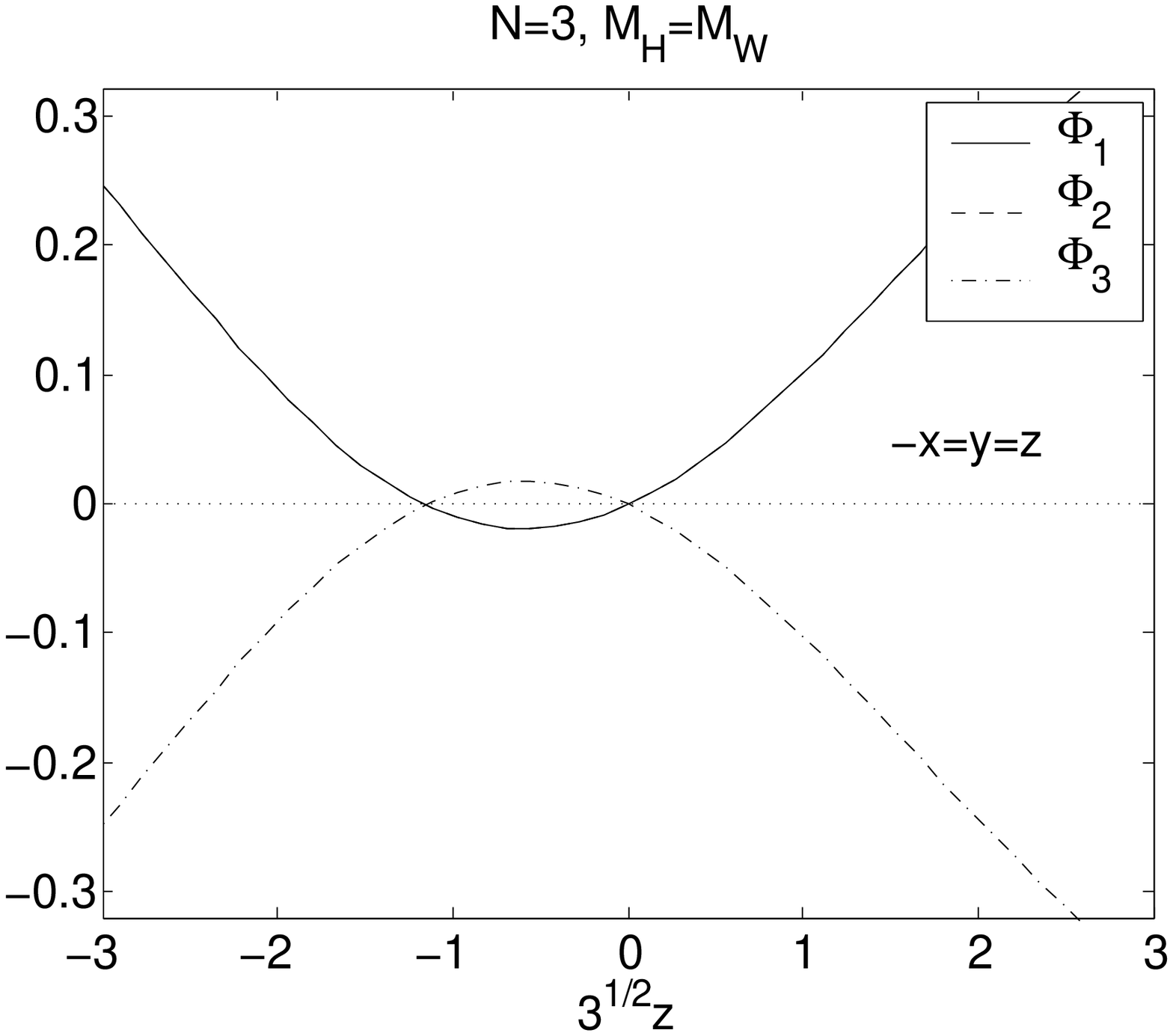}
 }
}
}
%\vspace{1.cm} }
\centerline{
\mbox{\epsfysize=6.0cm 
\epsffile{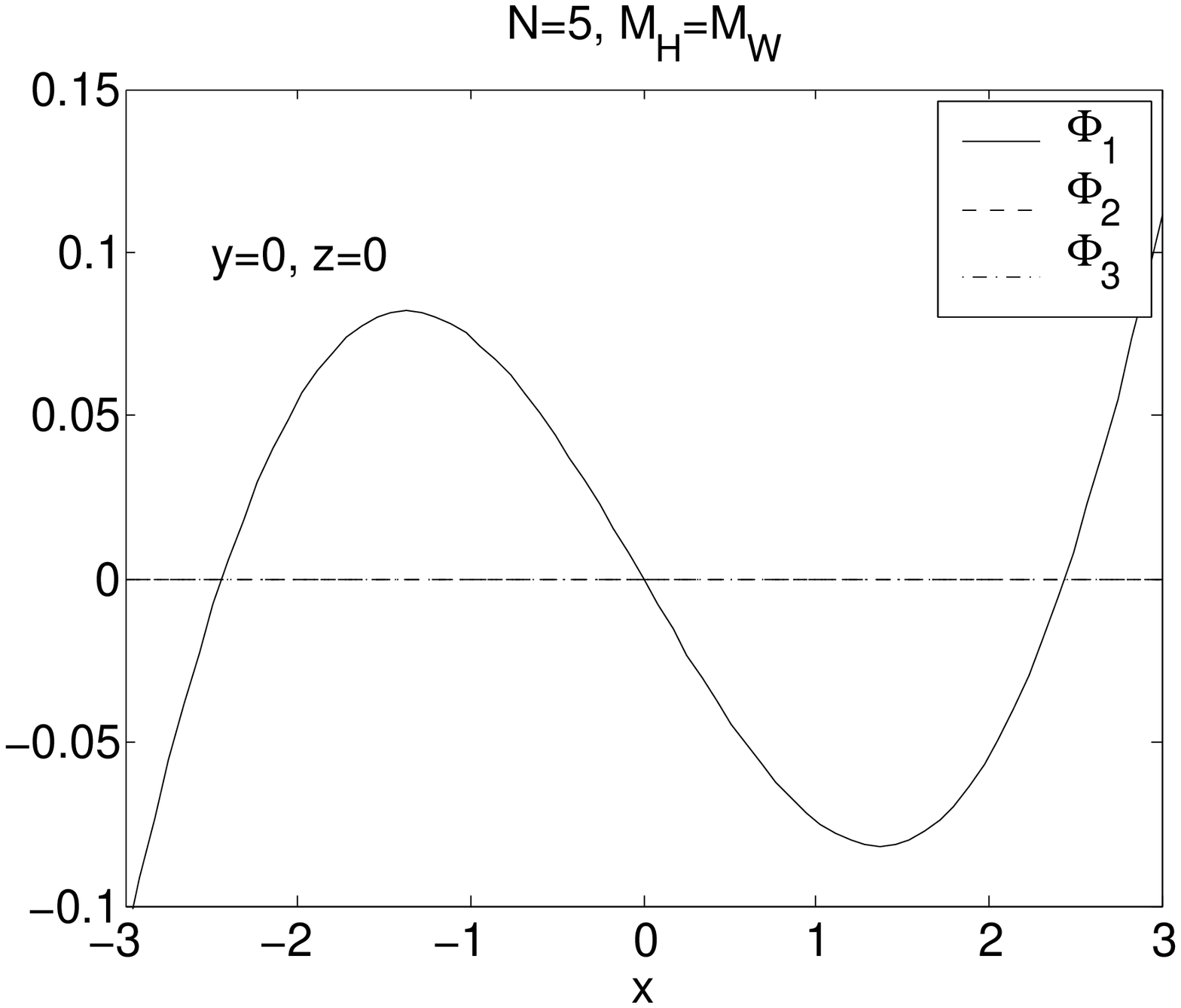}
 }
}

%\vspace{1.cm} 

The components of the Higgs field $\Phi_a$, $a=1,2,3$, are
shown (in units of $v/\sqrt{2}$)
for the tetrahedral sphaleron ($N=3$)
along the diagonals $x=-y=z$ (2a) and $-x=y=z$ (2b)
(in dimensionless coordinates $x$, $y$, $z$).
Along the diagonal $x=y=z$ all three components coincide with 
the component $\Phi_3$ of (2b), while along the diagonal $-x=-y=z$
$\Phi_1$ coincides with the component $\Phi_2$ of (2a),
and $\Phi_2=\Phi_3$ coincides with $\Phi_1$ of (2a).

For the octahedral sphaleron ($N=5$)
the components of the Higgs field $\Phi_a$ are shown
along the cartesian $x$-axis (2c).
Along the $y$-axis $\Phi_1=\Phi_3=0$ and $\Phi_2$ coincides
with $\Phi_1$ of (2c), while along the $z$-axis
$\Phi_1=\Phi_2=0$ and $\Phi_3$ coincides
with $\Phi_1$ of (2c).


\begin{thebibliography}{000}
\bibitem{foot1}
Note that for the bisphalerons of \cite{kb} $K^0$ does not 
vanish and contributes to the baryon number.
\end{thebibliography}
\end{document}